\documentclass{IEEEoj}
\usepackage{amsmath,amssymb,amsfonts}
\usepackage{algorithmic}
\usepackage{graphicx,color}
\usepackage{textcomp}
\usepackage{braket}
\usepackage{mathtools}
\usepackage{lipsum}
\usepackage{lscape}
\usepackage{multirow}
\usepackage{here}
\usepackage{bm}
\usepackage{cuted}
\usepackage{adjustbox}
\usepackage{cite}
\usepackage{tabularx}
\let\labelindent\relax
\usepackage{enumitem}
\usepackage{booktabs}
\usepackage{hyperref}
\hypersetup{%
 setpagesize=false,%
 bookmarks=true,%
 bookmarksdepth=tocdepth,%
 bookmarksnumbered=true,%
 colorlinks=true,%
 hidelinks,
 pdftitle={},%
 pdfsubject={},%
 pdfauthor={},%
 pdfkeywords={}}
\usepackage{makeidx}
\usepackage{setspace}

\usepackage{here}
\usepackage{subfigure}
\usepackage{amsthm}
\newtheorem{definition}{Definition}

\usepackage{array}
\usepackage{dsfont}
\usepackage[ruled,vlined]{algorithm2e}
\SetKwInput{KwInput}{Input}             
\SetKwInput{KwOutput}{Output}
\SetKwFor{Parfor}{par for}{}{}

\newcommand{\mb}[1]{\mathbf{#1}}

\newcommand{\prox}{{\rm prox}}
\newcommand{\argmax}{\mathop{\rm arg~max}\limits}
\newcommand{\argmin}{\mathop{\rm arg~min}\limits}

\def\BibTeX{{\rm B\kern-.05em{\sc i\kern-.025em b}\kern-.08em
    T\kern-.1667em\lower.7ex\hbox{E}\kern-.125emX}}
\AtBeginDocument{\definecolor{ojcolor}{cmyk}{0.93,0.59,0.15,0.02}}
\begin{document}
\receiveddate{XX Month, XXXX}
\reviseddate{XX Month, XXXX}
\accepteddate{XX Month, XXXX}
\publisheddate{XX Month, XXXX}
\currentdate{XX Month, XXXX}

\title{Dynamic Sensor Placement Based on Sampling Theory for Graph Signals}

\author{SAKI NOMURA$^1$, JUNYA HARA$^2$, HIROSHI HIGASHI$^{2}$, AND YUICHI TANAKA$^{2}$}
\affil{Department of Electrical Engineering and Computer Science, Tokyo University of Agriculture and Technology, Tokyo, 184--8588 Japan}
\affil{Graduate School of Engineering, Osaka University, Osaka, 565--0871 Japan}
\corresp{CORRESPONDING AUTHOR: SAKI NOMURA (e-mail: s\_nomura@msp-lab.org).}
\authornote{This work is supported in part by JSPS KAKENHI under Grant 23K26110 and 23K17461, and JST AdCORP under Grant JPMJKB2307.}
\markboth{DYNAMIC SENSOR PLACEMENT BASED ON GRAPH SAMPLING THEORY}{NOMURA \textit{et al.}}

\begin{abstract}
In this paper, we consider a sensor placement problem where sensors can move within a network over time.
Sensor placement problem aims to select $K$ sensor positions from $N$ candidates where $K < N$.
Most existing methods assume that sensor positions are static, i.e., they do not move,
however, many mobile sensors like drones, robots, and vehicles can change their positions over time.
Moreover, underlying measurement conditions could also be changed, which are difficult to cover with statically placed sensors.
We tackle the problem by allowing the sensors to change their positions in their neighbors on the network.
We dynamically determine the sensor positions based on graph signal sampling theory such that the non-observed signals on the network can be best recovered from the observations.
For signal recovery, the dictionary is learned from a pool of observed signals. It is also used for the sensor position selection.
In experiments, we validate the effectiveness of the proposed method via the mean squared error of the reconstructed signals.
The proposed dynamic sensor placement outperforms the existing static ones for both synthetic and real data.
\end{abstract}



\maketitle

\section{INTRODUCTION}
\label{sec:intro}
\IEEEPARstart{S}{ensor} networks are used in various engineering applications such as traffic, infrastructure, and facility monitoring systems \cite{tilak_infrastructure_2002, oliveira_wireless_2011, yi_methodology_2012,visalini_sensor_2019}.
They aim to efficiently collect information from distributed sensor measurements to enable real-time monitoring and analysis.
Sensing targets include many environmental and/or behavioral data like temperature and acceleration \cite{wang_reinforcement_2019, manohar_data-driven_2018}.

Sensor placement problem is one of the main research topics in sensor networks \cite{lin_near-optimal_2005, sakiyama_eigendecomposition-free_2019,jiang_sensor_2020}. Its purpose is to find $K$ sensor positions among $N$ candidates $(K < N)$ where sensors can be placed optimally so that the reconstruction error is minimized in some sense.
It has been extensively studied in machine learning and signal processing \cite{downey_optimal_2018, li_exploring_2021, sakiyama_eigendecomposition-free_2019, joshi_sensor_2008}.

Data on a network can be measured by sensors located at its nodes like wireless sensor networks, the Internet, and power grids \cite{oliveira_wireless_2011}.
Therefore, sensor placement on a network, i.e., graph, has been widely studied \cite{sakiyama_efficient_2016}.
In the graph-based sensor placement problem, nodes of the graph represent possible sensor positions, edges correspond to relationships among the sensors, and the signal value on a node is a measurement.

Many existing works of sensor placement problem assume the static sensor placement, i.e., sensors do not or cannot move \cite{li_efficient_2021,zhou_information-theoretic_2012,alonso_optimal_2004}.
However, the measurement conditions could be changed over time.
In this case, the selected sensors in the static positions do not reflect the current signal statistics. This may lead to the limitation of reconstruction qualities.

To overcome the limitation of the above-mentioned problem, in this paper, we propose a dynamic sensor placement where sensors can flexibly change their positions over time.

To realize the dynamic sensor placement, we sequentially learn the dictionary for signal reconstruction from a pool of observed signals on a network (i.e., graph signals) by utilizing sparse coding. 
We then dynamically determine the sensor positions at every time instance based on graph signal sampling theory \cite{tanaka_sampling_2020}, such that the non-observed graph signal can be best recovered from the sampled graph signal.

In experiments, we demonstrate that the proposed method outperforms existing sensor placement methods both in synthetic and real datasets.

\textit{Notation:}
Symbols used in this paper are summarized in Table \ref{tab:notation}.  
We consider a weighted undirected graph $\mathcal{G}=(\mathcal{V,E})$,
where $\mathcal{V}$ and $\mathcal{E}$ represent sets of nodes and edges, respectively. The number of nodes is $N = |\mathcal{V}|$ unless otherwise specified. The adjacency matrix of $\mathcal{G}$ is denoted by $\mb{W}$ where its $(m,n)$-element $\mb{W}_{mn}\geq 0$ is the edge weight between the $m$th and $n$th nodes; $\mb{W}_{mn}=0$ for unconnected nodes. The degree matrix $\mathbf{D}$ is defined as $\mathbf{D}=\text{diag}\,(d_{0},d_{1},\ldots,d_{N-1})$, where $d_{m}=\sum_n \mb{W}_{mn}$ is the $m$th diagonal element.
We use graph Laplacian $ \mathbf{L}\coloneqq \mathbf{D}-\mb{W}$ as a graph variation operator. A graph signal $\mb{x} \in \mathbb{R}^N$ is defined as a function $x:\mathcal{V}\rightarrow \mathbb{R}$.
Simply speaking, $x[n]$ is located on the node $n \in \mathcal{V}$ of $\mathcal{G}$.

The graph Fourier transform (GFT) of $\mb{x}$ is defined as
\begin{align}
\hat{x}(\lambda_i)=\braket{\mb{u}_{i},\mb{x}}=\sum_{n=0}^{N-1}u_{i}[n]x[n],
\end{align}
where $\mb{u}_i$ is the $i$th column of an orthonormal matrix $\mathbf{U}$.
It is obtained by the eigendecomposition of the graph Laplacian $\mathbf{L}=\mathbf{U\Lambda U}^\top$ with the eigenvalue matrix $\mb{\Lambda}=\mathrm{diag}\,(\lambda_0,\lambda_1,$ $\ldots,\lambda_{N-1})$. Without loss of generality, we assume the eigenvalues and corresponding eigenvectors are ordered in ascending order of the eigenvalues, i.e., $0=\lambda_0\leq\lambda_1\leq\cdots\leq\lambda_{N-1}$ and $\mathbf{U} = \begin{bmatrix}
    \mathbf{u}_0, \mathbf{u}_1, \dots, \mathbf{u}_{N-1}
\end{bmatrix}$. Since the eigenvector associated by lower $\lambda_i$ captures smoother components on the graph, we refer to $\lambda_i$ as a \textit{graph frequency}.

\section{GRAPH SIGNAL SAMPLING}\label{sec:graph_sampling}

In this section, we introduce graph signal sampling. We first review sampling theory on graphs and then describe a sampling set selection on graphs as presented in \cite{hara_sampling_2022}.

\subsection{Sampling Theory on Graphs}
\label{subsec:sampling_theory}

A key difference between the sampling of graph signals and that of standard signals stems from the underlying signal model \cite{tanaka_sampling_2020}. For example, smooth signals on a graph are characterized by small variations in signal values between adjacent nodes. Here, we assume that sensor measurements are implicitly governed by the underlying graph topology.

Suppose that a graph signal is generally represented by the following model \cite{tanaka_sampling_2020}:
\begin{align}
\mb{x}\coloneqq \mb{A}\mb{d},\label{eq:generator}
\end{align}
where $\mb{A}\in\mathbb{R}^{N\times M}\ (M\leq N)$ is a known generator matrix and $\mb{d}\in\mathbb{R}^{M}$ is a vector containing expansion coefficients.
We can also assume that the small number of atoms (i.e., columns in $\mathbf{A}$) essentially contributes to the signal $\mb{x}$.
$M$ can be set to an arbitrary number depending on the application. As explained below, examples of $M$ include bandwidth, the number of clusters or timeslots observed, etc.
The generator matrix $\mb{A}$ specifies the signal subspace $\mathcal{A}$, i.e., $\mathcal{A}=\mathcal{R}(\mb{A})$.

\begin{table}[!t]
\centering
\caption{Symbols used in this paper.}\label{tab:notation}
\begin{tabular}{c|c}
\hline
$N$ & Number of nodes\\
$K$ & Number of sensors\\
$T$ & Time duration (date length)\\
$D$ & Number of time slots \\
$t$ & Index of the time instance \\\hline
$\|\cdot\|$ & $\ell_2$ norm\\
$\|\cdot\|_1$ & $\ell_1$  norm\\
$\|\cdot\|_F$ & Frobenius norm\\
$y_i$ & $i$th element of $\mb{y}$\\
$\mb{Y}_{:i}$ & $i$th column of $\mb{Y}$\\
$\mb{Y}_{mn}$ & $(m,n)$-th element of $\mb{Y}$\\
$\mb{Y}_{\mathcal{XY}}$ & Submatrix of $\mb{Y}$ indexed by $\mathcal{X}$ and $\mathcal{Y}$\\
$\mb{Y}_{\mathcal{X}}$ & $\mb{Y}_{\mathcal{XX}}$\\
$\mathcal{X}^c$ & Complement set of $\mathcal{X}$\\
$\mathcal{R}(\mb{Y})$ & Range space of $\mb{Y}$\\\hline
\end{tabular}
\end{table}

The bandlimited (BL) model is one of the well-studied generators \cite{tanaka_sampling_2020,chen_discrete_2015,anis_efficient_2016,tsitsvero_signals_2016,puy_random_2018,wang_local_2016}, defined as
\begin{align}
    \mb{x}=\sum_{i=0}^{M-1}d_i\mb{u}_i=\mb{U}_{\mathcal{VM}}\mb{d},\label{eq:bandlimited_model}
\end{align}
where $\mb{U}_{\mathcal{VM}}\in\mathbb{R}^{N\times M}$ is the submatrix of $\mb{U}$ whose columns are specified by $\mathcal{M}=\{1,\ldots,M\}$. 
In this case, the generator matrix is $\mb{A} = \mb{U}_{\mathcal{VM}}$ where $M$ is the bandwidth.
     
Piecewise constant (PC) model \cite{chen_representations_2016} is another possible signal model and is defined as follows.
\begin{equation}
    \mb{x} = \sum_{i=0}^{M-1} d_i \mathbf{1}_{\mathcal{T}_i} = [\mathbf{1}_{\mathcal{T}_1}, \dots, \mathbf{1}_{\mathcal{T}_M}] \mb{d},\label{eq:PC}
\end{equation}
where $\mathcal{T}_i \, (i=1, \ldots, M)$ is the nonoverlapping subset of nodes for the $i$th cluster: $\mathbf{1}_{\mathcal{T}_i}$ is the cluster indicator vector: $[\mathbf{1}_{\mathcal{T}_i}]_n=1$ if $n\in\mathcal{T}_i$ and  $[\mathbf{1}_{\mathcal{T}_i}]_n=0$ otherwise \cite{chen_representations_2016}.
In this case, the generator matrix is $\mb{A} = [\mathbf{1}_{\mathcal{T}_1}, \dots, \mathbf{1}_{\mathcal{T}_M}]$ where $M$ is the number of clusters.

If we do not know the exact $\mb{A}$ but have its prior knowledge, it may be learned from the observations.
Let $\mb{X}\in\mathbb{R}^{N\times M}$ be a collection of $M$ graph signals.
Typically, $\mb{A}$ has been determined by $\mb{A}=\bm{\mathcal{U}}$, where $\bm{\mathcal{U}}$ is obtained by the SVD of the collection of observed signals $\mb{X}\coloneqq\bm{\mathcal{U}}\mb{\Sigma}\bm{\mathcal{V}}^\top$ \cite{jayaraman_data-driven_2020}, and $\bm{\mathcal{U}}$ and $\bm{\mathcal{V}}$ are unitary matrices. 

Let $\mb{S}^\top\in\mathbb{R}^{K\times N} (K\leq N)$ be an arbitrary linear sampling operator where columns of $\mb{S}$ are linearly independent.
Hereafter, we assume $K=M$ for simplicity, while we can easily extend the case of  $K < M$.
It specifies the sampling subspace $\mathcal{S}$, i.e., $\mathcal{S}=\mathcal{R}(\mb{S})$.
A sampled graph signal is represented by 
\begin{equation}
\bm{c}=\mb{S}^\top(\mb{x}+\mb{m}),
\label{eq:samp_noise}
\end{equation}
where $m[i]\sim\mathcal{N}(0,\sigma^2)$ is white Gaussian noise.
It is best recovered by the following transform \cite{eldar_sampling_2015,tanaka_generalized_2020,hara_graph_2023}:
\begin{align}
    \tilde{\mb{x}}
    =\mb{A}(\mb{S}^\top\mb{A})^\dagger\mb{c},\label{eq:recov_subspace}
\end{align}
where $\cdot^\dagger$ is the Moore-Penrose pseudoinverse.

If $\mathcal{A}$ and $\mathcal{S}$ together span $\mathbb{R}^N$ and only intersect at the origin, 
perfect recovery, i.e.,  $\tilde{\mb{x}} = \mb{x}$, is guaranteed. 
This is referred to as 
the \textit{direct sum} (DS) condition \cite{tanaka_sampling_2020}.
Note that the DS condition implies whether $\mb{S}^\top \mb{A} \mb{A}^\top \mb{S}$ is invertible, since $(\mb{S}^\top\mb{A})^\dagger = \mb{A}^\top \mb{S}(\mb{S}^\top \mb{A} \mb{A}^\top \mb{S})^{-1}$ holds if and only if the DS condition is satisfied.

From the perspective of the sensor placement, $\mb{S}^\top$ can be considered as a node-wise sampling operator. 
In the following, we introduce a design method of $\mb{S}^\top$ on $\mathcal{G}$ for an arbitrary generator.

\begin{figure*}[t!]
    \centering
    \includegraphics[keepaspectratio, width=0.8\linewidth]{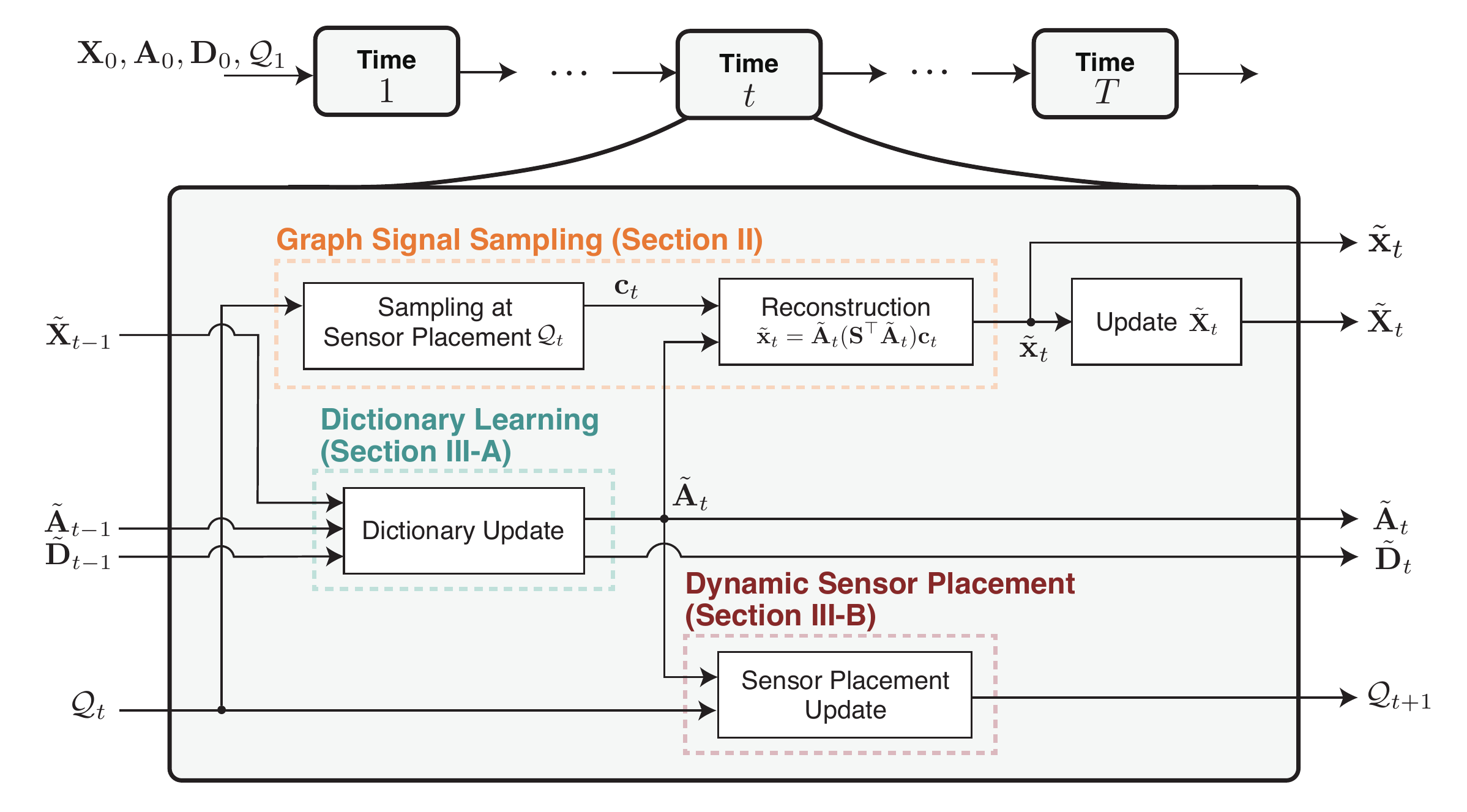}
    \caption{Overview of the proposed method.}
    \label{fig:proposed_flow}
\end{figure*}

\subsection{D-optimal Sampling Strategy}
\label{subsec:Sampling_Strategy}

We introduce a graph signal sampling method utilized for our dynamic sensor placement \cite{hara_graph_2023}.
Here, we define a node-wise sampling operator as follows:
\begin{definition}[Node-wise graph signal sampling]\label{def:node_samp_ope}
Let $\mb{I}_{\mathcal{MV}}\in\{0,1\}^{K\times N}$ be the submatrix of the identity matrix indexed by $\mathcal{M}\subset \mathcal{V}$ $(|\mathcal{M}|=K)$ and $\mathcal{V}$. The sampling operator is defined as
\begin{align}
\mb{S}^\top\coloneqq \mb{I}_{\mathcal{MV}}\mb{G},\label{eq:node_samp_ope}
\end{align}
where $\mb{G}\in\mathbb{R}^{N\times N}$ is an arbitrary graph filter. 
\end{definition}

While $\mb{G}$ is arbitrary, a typical selection of $\mb{G}$ is the identity matrix or a graph lowpass filter as that in classical sampling theory.
In a sensor placement perspective, $\mb{I}_{\mathcal{MV}}$ specifies the sensor positions on $\mathcal{G}$ and $\mb{G}$ can be viewed as a coverage area of sensors.
For example, we consider the following polynomial filter
\begin{align}
[\mb{G}]_{nm}=&\sum_{i=0}^{N-1}\sum_{j=0}^{J}\alpha_j\lambda^j u_i[n]u_i[m]\nonumber\\
=&\left[\mb{U}\left(\sum_{j=0}^{J}\alpha_j\mb{\Lambda}^j\right)\mb{U}^\top\right]_{nm},
\label{eq:phop_local_op}
\end{align}
where $\{\alpha_j\}$ are arbitrary coefficients. In graph signal processing, \eqref{eq:phop_local_op} is referred to as a $J$-hop localized filter\cite{shuman_vertex-frequency_2016}, whose nonzero response is limited within $J$-hop neighbors of the target node. This implies that the coverage area of all sensors is limited to $J$-hop.

Sensor placement problem is generally formulated as
\begin{align}
\mathcal{M}^*=\argmax_{\mathcal{M}\subset \mathcal{V}} e(\mathcal{M}),
\end{align}
where $e$ is a properly designed cost function. 

Typically, $e$ is designed based on the optimal experimental designs \cite{condat_primaldual_2013}, such as  A-, E-, and D-optimality. 
D-optimality is one of the popular experimental designs and is widely used in graph signal sampling.
The D-optimality is more appropriate than other optimality designs if we aim to recover graph signals from non-ideally sampled signals \cite{hara_graph_2023}.

In the D-optimal design, the following problem is considered as selecting an optimal sampling set for graph signals \cite{hara_graph_2023}:
\begin{align}
    \mathcal{M}^*=\mathop{\arg \max}_{\mathcal{M}\subset \mathcal{V}} \det (\mb{Z}_\mathcal{M}),\label{eq:detmax1}
\end{align}
where $\mb{Z}=\mb{G}\mb{A}\mb{A}^\top\mb{G}$ and thus $\mb{Z}_\mathcal{M} = \mb{I}_{\mathcal{MV}}\mb{G}\mb{A}\mb{A}^\top\mb{G}\mb{I}^{\top}_{\mathcal{MV}}$ (see Table~\ref{tab:notation} and Definition~\ref{def:node_samp_ope}). Recall that $\mb{S}^\top\mb{A}\mb{A}^\top\mb{S}$ is invertible when the DS condition is satisfied. Therefore, the maximization of $\det(\mb{Z}_\mathcal{M})=\det(\mb{S}^\top\mb{A}\mb{A}^\top\mb{S})=|\det(\mb{S}^\top\mb{A})|^2$ leads to the best recovery of graph signals.

The direct maximization of \eqref{eq:detmax1} is combinatorial and is practically intractable.
Therefore, a greedy method is applied to \eqref{eq:detmax1}.
Suppose that $\text{rank}(\mb{Z})\geq K$.  
In \cite{zhang_schur_2006}, \eqref{eq:detmax1} can be converted to a greedy selection as follows:
\begin{align}
y^*=&\mathop{\arg \max}_{y\in \mathcal{M}^c} \det (\mb{Z}_{\mathcal{M}\cup\{y\}})\nonumber\\
= &\mathop{\arg \max}_{y\in \mathcal{M}^c}\mb{Z}_{yy}-\mb{Z}_{y\mathcal{M}}(\mb{Z}_{\mathcal{M}})^{-1}\mb{Z}_{\mathcal{M}y}.\label{eq:maxdet2}
\end{align}
The selection is performed so that the best node $y^*$ is appended to the existing sampling set $\mathcal{M}$ one by one.

\section{DYNAMIC SENSOR PLACEMENT ON GRAPHS}
\label{sec:dynamic_sensor_placement}
In this section, we propose a dynamic sensor placement problem based on sampling theory on graphs. 
First, we derive online dictionary learning based on sparse coding. Second, we consider the control law of sensors.
Fig.~\ref{fig:proposed_flow} shows the overview of the proposed method.
The flow of the proposed method is outlined as follows:
\begin{enumerate}
    \item The dictionary is sequentially learned from previous observations (Section \ref{sec:dynamic_sensor_placement}-\ref{subsec:online_dictionary_learning}).
    \item Time-varying graph signals are sampled and reconstructed based on graph sampling theory (Section \ref{sec:graph_sampling}).
    \item Sensor positions are dynamically determined based on the dictionary (Section \ref{sec:dynamic_sensor_placement}-\ref{subsec:dynamic_sensor_placement}).
\end{enumerate}

\subsection{Online Dictionary Learning}
\label{subsec:online_dictionary_learning}
In sampling and reconstruction introduced in Section \ref{sec:graph_sampling}-\ref{subsec:sampling_theory}, $\mb{A}$ in \eqref{eq:generator} is assumed to be fixed.
Note that \eqref{eq:maxdet2} is regarded as a static sensor placement problem since the sensor positions are determined only once. In contrast, we consider the time-varying $\mb{A}_t$ in this paper where the subscript $t$ denotes the time instance (see Table~\ref{tab:notation}). In this setting, the optimal sensor positions can be changed according to $\mb{A}_t$. 

We infer a time-varying generator $\mb{A}_t$ from previous $D$ observations
\begin{equation}
    \mb{X}_{t-1}=[\mb{x}_{t-D},\dots,\mb{x}_{t-1}],
    \label{eq:Xt-1}
\end{equation}
in every time instance for dynamic sensor placement.

We assume the measurement model as follows:
\begin{equation}
    \mb{X}_t = f(\mb{X}_{t-1}),\label{eq:measure}
\end{equation}
where $f$ is a mapping from $\mb{X}_{t-1}$ to $\mb{X}_t$. That is, every observation is generated from signals in the previous time instance. 
Although the mapping of the time-evolution, $f$, is generally unknown, we suppose that $f$ is Lipschitzian, i.e., there exists some constant $L$ that satisfies the following inequality.
\begin{equation}
    \|f(\mb{X}_{t-1})-f(\mb{X}_{t})\|_F\leq L \|\mb{X}_{t-1}-\mb{X}_t\|_F.\label{eq:lipschitzian}
\end{equation}
By definition of $f$, \eqref{eq:lipschitzian} can be transformed into
\begin{equation}
    \|\mb{X}_{t}-\mb{X}_{t+1}\|_F\leq L \|\mb{X}_{t-1}-\mb{X}_t\|_F.
    \label{eq:lipschitzian_2}
\end{equation}
In this paper, we assume a non-expansive mapping in \eqref{eq:measure}, i.e., $f$ with the Lipschizian constant $L\leq1$, to ensure that the signals are reasonably smooth over time. This assumption is essential for learning the dictionary stably, while we do not explicitly estimate $f$ in the following part.

Next, we consider estimating the generator $\mb{A}_t$ as a dictionary from (the estimated) $\tilde{\mb{X}}_{t-1}$ in \eqref{eq:Xt-1} based on the assumption in \eqref{eq:lipschitzian_2}, where $\tilde{\cdot}$ corresponds to the estimated vector/matrix.

Since the exact value of the number of atoms, $M$ in \eqref{eq:bandlimited_model} and \eqref{eq:PC} may not be known in general, herein, we suppose $M \ll D$, and denote a dictionary and expansion coefficient by $\mb{A}_t\in \mathbb{R}^{N \times D}$ and $\mb{d}_t\in\mathbb{R}^D$, respectively. 
Therefore, we seek the optimal $\mb{A}_t$ such that the number in nonzero values of $\mb{d}_t$ is close to $M$, i.e., $\mb{d}_t$ is sparse. This setting is similar to well-studied dictionary learning problems \cite{aharon_k-svd_2006, engan_method_1999}.

Let $\mb{D}_t = [\mb{d}_{t-D+1},\dots,\mb{d}_{t}] \in\mathbb{R}^{D\times D}$ be the collection of expansion coefficients from time $t-D+1$ to $t$. 
Then, we can express $\mb{X}_t=\mb{A}_t\mb{D}_t$ (see \eqref{eq:generator}). 

Here, we assume that the graph $\mathcal{G}$, training data $\mb{X}_0$, initial generator matrix $\mb{A}_0$, and expansion coefficients matrix $\mb{D}_0$ are given, where $\mb{A}_0$ and $\mb{D}_0$ are arbitrary matrices.
Following from \eqref{eq:lipschitzian_2}, we can suppose that $\mb{X}_{t}$ is sufficiently close to $\mb{X}_{t-1}$. 
As a result, we formulate the following problem:
\begin{align}
    \argmin_{\tilde{\mb{A}}_t,\tilde{\mb{D}}_t} \|\tilde{\mb{X}}_{t-1}-\tilde{\mb{A}}_t \tilde{\mb{D}}_t\|_F^2+\mu\|\tilde{\mb{D}}_t\|_1 + \eta\|\tilde{\mb{A}}_{t} - \tilde{\mb{A}}_{t-1}\|_F^2,
    \label{eq:online_dic_learn}
\end{align}
where $\tilde{\mb{X}}_{t-1}=\tilde{\mb{A}}_{t-1}\tilde{\mb{D}}_{t-1}$, and $\mu$ and $\eta$ are the parameters.
The first term in \eqref{eq:online_dic_learn} is the data fidelity term for $\tilde{\mb{A}}_t$. The second and third terms are for the regularization controlling the sparsity of $\tilde{\mb{D}}_t$ and the temporal variation of $\tilde{\mb{A}}_t$, respectively.

Note that we need to solve \eqref{eq:online_dic_learn} with respect to two variables $\tilde{\mb{A}}_t$ and $\tilde{\mb{D}}_t$, which jointly form a nonconvex optimization.
In this paper, we divide \eqref{eq:online_dic_learn} into two independent subproblems with respect to $\tilde{\mb{A}}_t$ and $\tilde{\mb{D}}_t$, and solve them alternately, similar to the method in \cite{engan_method_1999}.

First, we solve \eqref{eq:online_dic_learn} with respect to $\tilde{\mb{A}}_t$ by fixing $\tilde{\mb{D}}_{t}$ as $\tilde{\mb{D}}_{t-1}$.
In this case, we easily obtain the closed-form solution for the dictionary as follows:
\begin{align}
    \tilde{\mb{A}}_{t}
    &=\argmin_{\tilde{\mb{A}}_t}\|\tilde{\mb{X}}_{t-1}-\tilde{\mb{A}}_t\tilde{\mb{D}}_{t-1}\|_F^2 + \eta\|\tilde{\mb{A}}_{t} - \tilde{\mb{A}}_{t-1}\|_F^2 \nonumber\\
    &= (\eta \tilde{\mb{A}}_{t-1} + \tilde{\mb{X}}_{t-1}\tilde{\mb{D}}_{t-1}^\top) (\eta \mb{I}_D + \tilde{\mb{D}}_{t-1}\tilde{\mb{D}}_{t-1}^\top)^{-1}.
    \label{eq:gen_update}
\end{align}

Second, we update $\tilde{\mb{D}}_t$ in \eqref{eq:online_dic_learn} by using $\tilde{\mb{A}}_t$ in \eqref{eq:gen_update}, i.e., 
\begin{align}
    \tilde{\mb{D}}_t = \argmin_{\tilde{\mb{D}}_t} \|\tilde{\mb{X}}_{t-1}-\tilde{\mb{A}}_t \tilde{\mb{D}}_t\|_F^2+\mu\|\tilde{\mb{D}}_t\|_1.\label{eq:online_dic_learn_d}
\end{align}
To solve \eqref{eq:online_dic_learn_d}, we utilize the proximal gradient method\cite{passty_ergodic_1979, tseng_applications_1991, combettes_signal_2005}, whose problem is given by the following form.
\begin{align}
    \tilde{\mb{D}}_t = \argmin_{\tilde{\mb{D}}_t} g(\tilde{\mb{D}}_t)+ h(\tilde{\mb{D}}_t),\label{eq:prox_grad}
\end{align}
where $g$ is a differentiable function and $h$ is a proximable function\cite{condat_primaldual_2013}. Note that \eqref{eq:prox_grad} is set to the entire problem being convex. 
As a result, the optimal solution is obtained by the following update rule.
\begin{align}
    \tilde{\mb{D}}_{t,n+1}=\prox_{\gamma h}(\tilde{\mb{D}}_{t,n}-\gamma \nabla g(\tilde{\mb{D}}_{t,n})),\label{eq:prox_grad_update}
\end{align}
where $n+1$ is the number of iterations, $\gamma$ is the step size, $\nabla g$ is the gradient of $g$, and $\prox_{\gamma h}$ is defined as
\begin{align}
    \prox_{\gamma h}(\mb{D})
    &=\argmin_{\mb{C}}\frac{1}{2}\|\mb{C}-\mb{D}\|^2_F+ \gamma\mu\|\mb{C}\|_1 \nonumber\\
    &= S_{\gamma\mu}(\mb{D}),
    \label{eq:prox_D}
\end{align}
where $S_{\gamma \mu}$ is the soft thresholding operator defined as follows: 
\begin{equation}
    [S_{\gamma\mu}(\mb{D})]_{ij}= \begin{cases}\mb{D}_{ij}-\gamma\mu, & \mb{D}_{ij} \geq \gamma\mu \\ 0, & |\mb{D}_{ij}|<\gamma\mu \\ \mb{D}_{ij}+\gamma\mu, & \mb{D}_{ij} \leq-\gamma\mu.\end{cases}
\end{equation}
By applying  \eqref{eq:prox_D} to \eqref{eq:prox_grad_update}, we have the following iteration with ISTA \cite{beck_fast_2009}.
\begin{align}
    \tilde{\mb{D}}_{t,n+1} 
        &= S_{\gamma \mu}(\tilde{\mb{D}}_{t,n} - \gamma \nabla g(\tilde{\mb{D}}_{t,n})) \nonumber \\
        &= S_{\gamma \mu}\{\tilde{\mb{D}}_{t,n} - 2\gamma \tilde{\mb{A}}_t^\top (\tilde{\mb{A}}_t \tilde{\mb{D}}_{t,n} - \tilde{\mb{X}}_{t-1})\},
        \label{eq:prox_itr}
\end{align}
We set the step size as $\gamma \leq 1/ {\lambda_\text{max}(\tilde{\mb{A}}_t^\top \tilde{\mb{A}}_t)}$ according to the convergence condition of ISTA \cite{beck_fast_2009}. Therefore, the proposed algorithm is guaranteed to converge.
We iterate \eqref{eq:prox_itr} until $\|\tilde{\mb{D}}_{t,n+1}-\tilde{\mb{D}}_{t,n}\|_F^2<\epsilon$ where $\epsilon$ is a small constant.
The optimization of \eqref{eq:online_dic_learn} is summarized in Algorithm \ref{algo:online_dic_learn}. 

\begin{algorithm}[h!]
    \DontPrintSemicolon
    \setlength{\abovedisplayskip}{0pt}
    \setlength{\belowdisplayskip}{0pt}
      \KwInput{$\tilde{\mb{X}}_{t-1}, \tilde{\mb{A}}_{t-1}, \tilde{\mb{D}}_{t-1}, t$}
      \eIf{$t=1$}{
          $\tilde{\mb{A}}_1 = \mb{A}_0$
    }{
        $\tilde{\mb{A}}_{t}
        = (\eta \tilde{\mb{A}}_{t-1} + \tilde{\mb{X}}_{t-1}\tilde{\mb{D}}_{t-1}^\top) (\eta \mb{I}_D + \tilde{\mb{D}}_{t-1}\tilde{\mb{D}}_{t-1}^\top)^{-1}$
    }
    $\tilde{\mb{D}}_t = \tilde{\mb{D}}_{t-1}$ \\
      \While{$\|\tilde{\mb{D}}_{t,n}-\tilde{\mb{D}}_{t,n-1}\|_F^2\geq\epsilon$}{
        $\tilde{\mb{D}}_{t,n+1} 
        = S_{\gamma \mu}\{\tilde{\mb{D}}_{t,n} - 2\gamma \tilde{\mb{A}}_t^\top (\tilde{\mb{A}}_t \tilde{\mb{D}}_{t,n} - \tilde{\mb{X}}_{t-1})\}$ \\
        $n \leftarrow n+1$
      }
      \KwOutput{$\tilde{\mb{A}}_t$}
    \caption{Online Dictionary Learning}\label{algo:online_dic_learn}
\end{algorithm}

In the following, we formulate the dynamic sensor placement based on the learned dictionary.
The control law of sensors is also derived based on the sampling strategy introduced in \eqref{eq:maxdet2}.

\subsection{Dynamic Sensor Placement}
\label{subsec:dynamic_sensor_placement}
For brevity, we define the following matrix:
\begin{align}
\mb{N}_t\coloneqq \tilde{\mb{A}}_t^\top\mb{G}.\label{eq:sensor_proj_mat}
\end{align}
Note that we now assume that $\tilde{\mb{A}}_t$ is given.
With \eqref{eq:sensor_proj_mat}, \eqref{eq:maxdet2} can be rewritten by a dynamic form as\footnote{For notation simplicity, we omit $\cdot_t$ in $\mb{N}_t$ hereafter.}
\begin{align}
\begin{split}
y_t^* = 
&\argmax_{y\in\mathcal{M}^c}\mb{N}_{:y}^\top\mb{N}_{:y}-\mb{N}_{:y}^\top\mb{N}_{:\mathcal{M}}(\mb{N}_{:\mathcal{M}}^\top\mb{N}_{:\mathcal{M}})^{-1}\mb{N}_{:\mathcal{M}}^\top\mb{N}_{:y} \\
= &\argmax_{y\in\mathcal{M}^c}\|\bm{\nu}_y\|^2-\|\mb{P}_{\mathcal{R}(\mb{N}_{:\mathcal{M}})}\bm{\nu}_y\|^2,\label{eq:prop_stat}
\end{split}
\end{align}
where $\bm{\nu}_y=\mb{N}_{:y}$ and $\mb{P}_{\mathcal{R}(\mb{Q})}$ is the orthogonal projection onto $\mathcal{R}(\mb{Q})$ (both symbols are defined in Table \ref{tab:notation}). Since $\mb{N}$ can temporally vary, the optimal solution in \eqref{eq:prop_stat} can also change in every time instance.

Note that \eqref{eq:prop_stat} is based on a greedy selection introduced in Section \ref{sec:graph_sampling}-\ref{subsec:Sampling_Strategy}.
However, in practice, this optimization may not be efficient for large graphs: The control law with \eqref{eq:prop_stat} implies that we need a large computational burden 
if all sensors' positions are sequentially determined at every time instance.
This results in a delay for sensor relocations which should be alleviated.
In this paper, instead, we consider selecting the sensor positions independently, i.e., \eqref{eq:prop_stat} is solved for a sensor independently of the other ones. In the following, we rewrite \eqref{eq:prop_stat} as a distributed optimization. 

Here, we denote the position of the $i$th sensor at the $t$th time instance by $q_{i,t}$. We define the $i$th Voronoi region in the graph $\mathcal{G}$ as follows:

\begin{figure}[t!]
  \centering
  \includegraphics[keepaspectratio, scale=0.3]
       {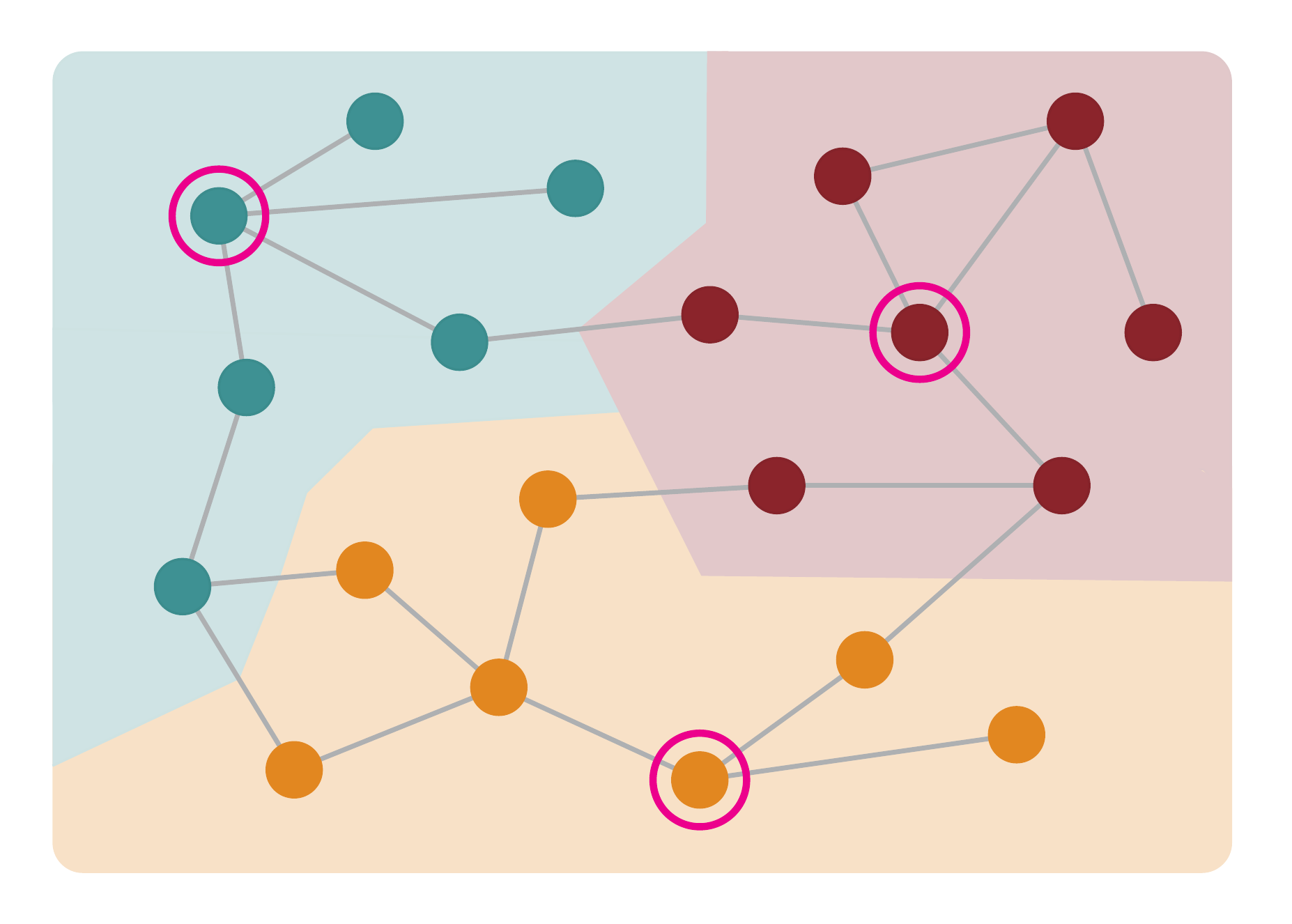}
  \caption{
    Voronoi diagram on a graph. Nodes with circles represent the current sensor positions. 
    Each colored area represents the Voronoi region corresponding to each sensor.
  }
  \label{fig:voronoi}
\end{figure}

\begin{algorithm}[t!]
\DontPrintSemicolon
\setlength{\abovedisplayskip}{0pt}
\setlength{\belowdisplayskip}{0pt}
  \KwInput{Sensing matrix $\mb{G}$, time duration $T$, initial data slot $\mb{X}_0$, initial dictionary $\mb{A}_0$, initial expansion coefficients matrix $\mb{D}_0$, initial positions of sensors $\mathcal{Q}_1$}
  \For{$t=1,\ldots,T$}{
    Update the dictionary $\tilde{\mb{A}}_t$ with Algorithm \ref{algo:online_dic_learn}\\
    Sample the current signal $\mb{x}_t$ at the positions $\mathcal{Q}_t$ in \eqref{eq:samp_noise} \\
    Reconstruct the current signal $\tilde{\mb{x}}_t$ in \eqref{eq:recov_subspace} \\
    Update the data matrix $\tilde{\mb{X}}_t$ by inserting $\tilde{\mb{x}}_t$\\
    Calculate the Voronoi regions $\{\mathcal{W}_{i,t}\}_{i=1,\ldots,K}$\\
    $\mb{N}_t\leftarrow \tilde{\mb{A}}^\top_t\mb{G}$\\
    \Parfor{$i=1,\ldots,K$}{
      Calculate $\mb{P}_{\mathcal{R}(\mb{N}_{:\mathcal{M}_{i,t}})}$\\
      Calculate the $P$ hop neighbors $\{\mathcal{N}^P_{i,t}\}_{i=1,\ldots,K}$\\
      $q^*_{i,t+1}=\argmax_{y\in\mathcal{W}_{i,t}\cap \mathcal{N}_{i,t}^{P} }\|\bm{\nu}_y\|^2-\|\mb{P}_{\mathcal{R}(\mb{N}_{:\mathcal{M}_{i,t}})}\bm{\nu}_y\|^2$\\
    }
    $\mathcal{Q}_{t+1}\leftarrow \{q^*_{i,t+1}\}_{i=1,\ldots,K}$\\
  }
  \KwOutput{Sensor positions $\{\mathcal{Q}_t\}_{t=1,\ldots,T+1}$}
\caption{Dynamic Sensor placement}\label{algo:online_sensor_placement}
\end{algorithm}

\begin{align}
\mathcal{W}_{i,t}=\{v\in\mathcal{V}\mid d(v,q_{i,t})<d(v,q_{j,t}),\forall q_j\in\mathcal{Q}_t\},\label{eq:voronoi_region}
\end{align}
where $d(v,q_{i,t})$ is the shortest distance between $v$ and $q_{i,t}$, and  $\mathcal{Q}_t=\{q_{j,t}\}_{j=1,\ldots,K}$ denotes the set of all sensor positions at $t$. 
Note that, in each $\mathcal{W}_{i,t}$, only one sensor is contained: We illustrate its example in Fig. \ref{fig:voronoi}.
The Voronoi regions are determined by $k$-means clustering, which is robust to perturbations to some extent depending on the topology \cite{balcan_k_2020}. Detailed analysis of robustness is left for future work.

By utilizing \eqref{eq:voronoi_region}, we can rewrite \eqref{eq:prop_stat} for seeking the best position of a sensor within its Voronoi region $\mathcal{W}_{i,t}$ as follows:

\begin{align}
q^*_{i,t+1}=\argmax_{y\in\mathcal{W}_{i,t}}\|\bm{\nu}_y\|^2-\|\mb{P}_{\mathcal{R}(\mb{N}_{:\mathcal{M}_{i,t}})}\bm{\nu}_y\|^2.\label{eq:prop_dynamic1}
\end{align}
Since \eqref{eq:prop_dynamic1} depends only on the Voronoi region of a sensor, all sensor selections in \eqref{eq:prop_dynamic1} are independent of each other. Therefore, we can select sensors by \eqref{eq:prop_dynamic1} in a distributed fashion. 

When we assume mobile sensors, their movable areas could be restricted to their neighbors.
To reflect this, we can further impose a constraint on the Voronoi region where the movable nodes are restricted to
the $P$-hop neighbors of the target node.
The $P$-hop neighbor of the $i$th sensor (node) is defined as
\begin{align}
\mathcal{N}_{i,t}^P=\{v\in\mathcal{V}\mid d(v,q_{i,t})\leq P\}.
\end{align}
As a result, the proposed control law of sensors in \eqref{eq:prop_dynamic1} is further modified as follows:
\begin{align}
    q^*_{i,t+1}=\argmax_{y\in\mathcal{W}_{i,t}\cap \mathcal{N}_{i,t}^{P}} \Psi(y),\label{eq:prop_dynamic2}
\end{align}
where $\Psi(y) := \|\bm{\nu}_y\|^2-\|\mb{P}_{\mathcal{R}(\mb{N}_{:\mathcal{M}_{i,t}})}\bm{\nu}_y\|^2$.
Algorithm \ref{algo:online_sensor_placement} summarizes the proposed dynamic sensor placement method\footnote{We utilize the \texttt{MATLAB} notation where we denote the parallel processing in Algorithm \ref{algo:online_sensor_placement} as \textbf{par for}.}.

As mentioned in Section~\ref{sec:graph_sampling}-\ref{subsec:Sampling_Strategy}, we utilize the D-optimality design for sensor selection, which is a monotone submodular function. It is widely known that greedy sensor selection with monotone submodular functions yields a nearly optimal solution \cite{clark_greedy_2019}.

Note that we consider a distributed sensor placement problem, which is an approximation of the greedy (centralized) one in some sense. 
We address the gap between the distributed and centralized solutions.
Here, we denote the set of sensor positions obtained in a distributed manner by $\{q_i^*\}_{i=1,\ldots,K}$ and those in a centralized manner by $\{r_i^*\}_{i=1,\ldots,K}$.
From \cite[Theorem 4.1]{mirzasoleiman_distributed_2013}, we have $\sum_{i=1}^K \Psi (r_i^*) \ge \sum_{i=1}^K \Psi (q_i^*) \ge \frac{1}{K}\sum_{i=1}^K \Psi (r_i^*)$.

Note that our main contribution is to formulate dynamic sensor placement as graph signal sampling.
We may be able to use other methods, such as those described in \cite{zhang_sampling_2021} and \cite{mairal_online_2009}, as internal algorithms for dictionary learning.

\subsection{Computational Complexity}
We now address the computational complexity of the proposed sensor placement method.
Determining the sensor placement in \eqref{eq:maxdet2} requires $O(K^3+NK^2)$ computational complexity, where $O(K^3)$ complexity costs for computing the inverse in \eqref{eq:maxdet2}, and $O(NK^2)$ complexity requires for the greedy search of the optimal nodes.
While \eqref{eq:maxdet2} determines the positions from the whole network \cite{hara_sampling_2022}, our algorithm performs it within the Voronoi regions in the distributed manner.
As a result, the computational complexity of our algorithm results in $O(K(N+|\mathcal{E}|)+K^3+\max_i\{|\mathcal{W}_{i,t}|\})$ at most, where $O(K(N+|\mathcal{E}|))$ complexity requires for the division of Voroinoi regions and $O(\max_i\{|\mathcal{W}_{i,t}|\})$ complexity is needed for the node selection in the Voronoi reigions in the worst case.
By comparing $O(NK^2)$ and $O(K(N+|\mathcal{E}|)+\max_i\{|\mathcal{W}_{i,t}|\})$ complexities, the latter is usually lower as long as the graph is not very dense.

\section{EXPERIMENTS}
\label{sec:experiments}
In this section, we compare reconstruction errors of time-varying graph signals measured by selected positions of sensors.
We perform recovery experiments for synthetic and real datasets.

We suppose that the graph $\mathcal{G}$ is given a priori. The nodes represent the candidates of sensor positions, and the edges indicate the paths where sensors can move. Sampling is performed at every time instance. After sampling, the reconstructed signal at $t$, i.e., $\tilde{\mb{x}}_t$, is stored as the latest signal collection $\tilde{\mb{X}}_{t}$, which is used for the dictionary/sensor positions update.
Simultaneously, the oldest signal, $\tilde{\mb{x}}_{t-D}$, is discarded from $\tilde{\mb{X}}_t$.

\subsection{Synthetic Graph Signals}
\label{subsec:synthetic}
In this subsection, we perform dynamic sensor placement for synthetic data.

\subsubsection{Setup}
We compare the signal reconstruction accuracy of the proposed method with those of the following existing methods.

\begin{enumerate}
\renewcommand{\labelenumi}{\arabic{enumi})}
    \item \textit{Static1}: Static sensor placement with the fixed dictionary learned by the SVD only once at $t = 0$ \cite{jayaraman_data-driven_2020}. This method does not change the dictionary over time.
    \item \textit{Static2}: Static sensor placement (similar to Static1) with the non-fixed dictionary learned by the SVD at every time instance.    
\end{enumerate}
Since the SVD-based dictionary learning is still widely-used in many sensor placement applications\cite{manohar_data-driven_2018}, Static1 and Static2 are possible baseline methods.

We also compare the proposed method to the extreme case of the proposed method, i.e., $P=\infty$.
In this case, sensors can move anywhere in Voronoi region regardless of the current sensor positions.

\begin{figure*}[!t]
    \centering
    \includegraphics[keepaspectratio, width=0.8\hsize]{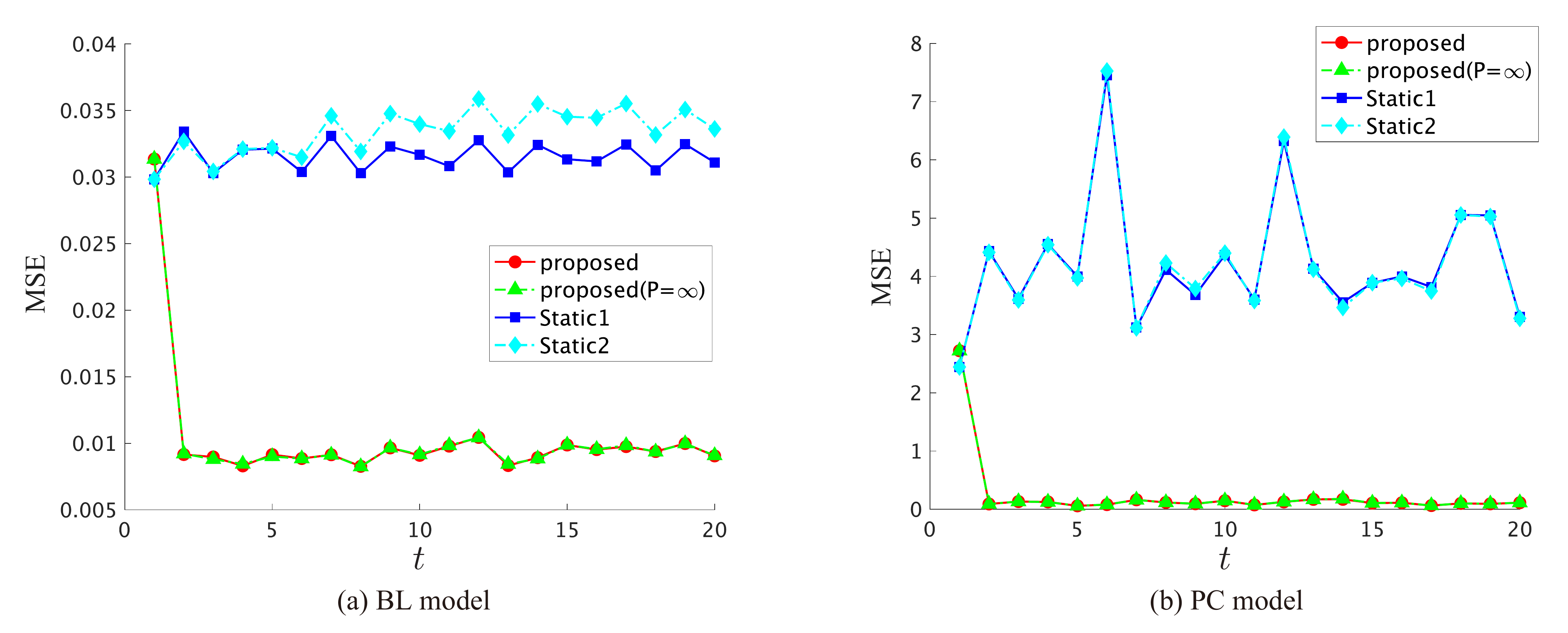}\caption{MSE comparison of the reconstructed synthetic data for the noisy case.}
    \label{fig:noisy}
\end{figure*}

\begin{figure*}[h]
    \centering\includegraphics[keepaspectratio, width=0.8\hsize]{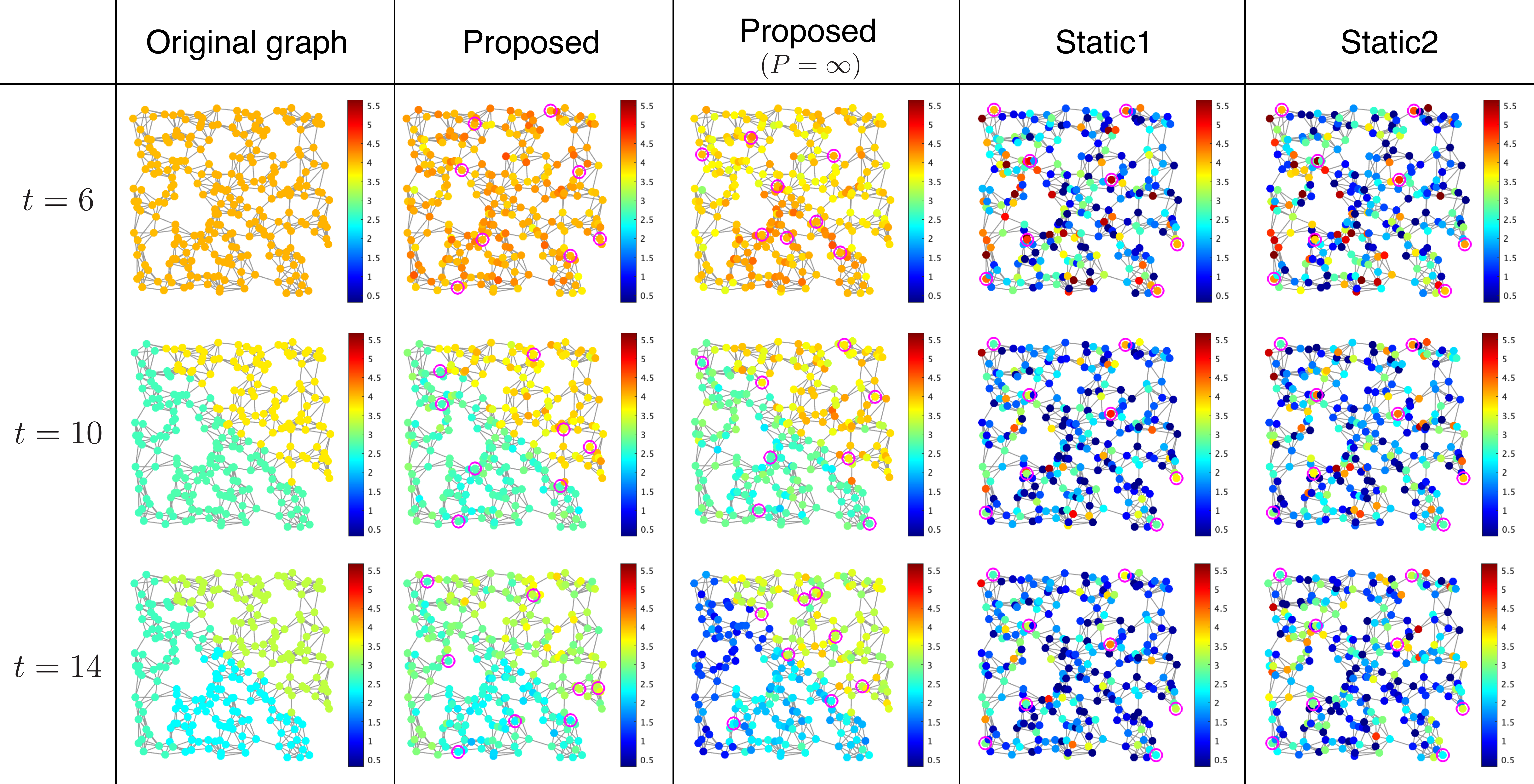}\caption{Reconstructed graph signals for PC signals at time $t=6,10,14$. Circled nodes represent selected sensor positions.}
    \label{fig:reconst}
\end{figure*}

In the synthetic case, the recovery experiment is performed for signals on a random sensor graph with $N=256$. The number of sensors is set to $K=8$. 
We consider the following graph signal models:
\begin{description}
    \item[Bandlimited (BL) model]\mbox{}\\
    The BL model is defined in \eqref{eq:bandlimited_model}.  We set to $M=N/16$\footnote{We set $K < M$ since perfect recovery is always guaranteed for $K \geq M$.}.
    We generate expansion coefficients $\mb{d}_t$ as follows:
    \begin{align}
    \mb{d}_t=
    \begin{bmatrix}
    \frac{1}{t} \times \sin(10t+\phi_i) +1\\
    \vdots\\
    \frac{1}{t} \times \sin(10t+\phi_M) +1
    \end{bmatrix},
    \end{align}
    where $\phi_i=\frac{\pi}{M}i$. In this model, graph signals temporally oscillate with the fixed bandwidth, and their oscillation amplitudes decay over time. It would mimic a possible physical data observation model.
    
    \item[Piecewise-constant (PC) model]\mbox{}\\ The PC model is defined in \eqref{eq:PC}. The number of clusters is set to $M=3$. We randomly separate nodes into $M$ clusters so that nodes in a cluster are connected. We generate $\mb{d}_t$ as follows:
    \begin{align}
    \mb{d}_t=
    \begin{bmatrix}
        3\{\exp({-\frac{t}{25}})\sin(t+\frac{\pi}{3})+1\}\\
        3\{\exp({-\frac{t}{25}})\sin(2t)+1\}\\
        3\{\exp({-\frac{t}{25}})\sin(3t-\frac{\pi}{3})+1\}
    \end{bmatrix}.
    \end{align}
    In this model, signal values in a cluster change simultaneously (but the temporal frequencies are different for different clusters).
    It can be a possible model for clustered sensor networks.
\end{description}
For both graph signal models, we generate the graph signals $\mb{x}_t$ for durations of $T_{\text{train}}=20$ for training and $T_{\text{test}}=20$ for test, respectively. The number of time slots is set to $D = 20$.

The initial data slot $\mb{X}_0$ used for the dictionary learning in \eqref{eq:online_dic_learn} is set to $\mb{X}_0 = \mb{X}_{\text{train}} + \mb{M}$, where $\mb{X}_{\text{train}}$ is the collection of original signals for training, and $\mb{M}_{i,j}\sim\mathcal{N}(0,0.1)$ is white Gaussian noise.
The initial expansion coefficients and dictionary can be arbitrary as mentioned in Section~\ref{sec:dynamic_sensor_placement}-\ref{subsec:online_dictionary_learning}.
In the experiment, however, we used an SVD-based dictionary for a fair comparison to alternative methods.
Therefore, $\mb{A}_{0}=\bm{\mathcal{U}}$ at $t=0$, where $\bm{\mathcal{U}}$ is obtained by the SVD of the known measurement data, i.e., $\mb{X}_{0}=\bm{\mathcal{U}}\mb{\Sigma}\bm{\mathcal{V}}^\top$. The initial value of $\mb{D}_t$ is also set to $\mb{D}_0=\mathbf{11}^\top$.
We set the temporal sampling period $T_s=\frac{\pi}{30}$ to avoid aliasing.
The parameters of the dictionary learning are experimentally set to $(\gamma,\eta,\mu)=(10^{-4}, 3, 1)$, and $P=1$ is used for the sensor movable area where we assume a sensor can only be moved to its one-hop neighbor based on reasonable sensor mobility.

We assume that $\mb{G}=\mb{U}\zeta(\boldsymbol{\Lambda})\mb{U}^\top$ is a graph low-pass filter. We define $\zeta(\boldsymbol{\Lambda}) =\operatorname{diag}\left(\zeta\left(\lambda_0\right), \zeta\left(\lambda_1\right), \ldots, \zeta\left(\lambda_{N-1}\right)\right)$ as follows:
\begin{equation}
    \zeta(\lambda_i)=\cos\left(\frac{\pi}{2}\cdot\frac{\lambda_i}{\lambda_\text{max}}\right). \label{eq:graph_filter}
\end{equation}

We apply the Chebyshev polynomial approximation \cite{stankovic_spectral_2019} to $\mathbf{G}$. This works as a $J$-hop localized filter. We set the order of approximation to $J=20$.

Recovery accuracy is measured in noisy scenarios where white Gaussian noise $\mb{m}_i \sim \mathcal{N} (0, 0.1)$ is added to the signals (see \eqref{eq:samp_noise}).
For all methods, the initial position of sensors is determined by a state-of-the-art sampling set selection method on graphs \cite{sakiyama_eigendecomposition-free_2019}. 
We calculate the averaged mean squared errors (MSEs) for 50 independent runs.

\subsubsection{Results}
Fig.~\ref{fig:noisy} shows the experimental results for the noisy case.
We also show examples of the original and reconstructed graph signals of PC models in Fig.~\ref{fig:reconst}. 

Fig.~\ref{fig:noisy} indicates that the proposed method shows consistently (and significantly) smaller MSEs than the other methods in all time instances, and the MSE of the proposed method gradually decreases over time.
This validates that the proposed method can adapt to the change in measurement conditions, while the static methods fail to do so.

In this experiment, $P=1$ and $\infty$ do not result in a large difference.
However, as shown in Fig.~\ref{fig:reconst} at $t=14$, we observe that $P=\infty$ produces a larger error than $P=1$ because several sensor locations are very close to each other.
A possible reason for this is that 
the sensor positions were biased toward the right side.

\subsection{Real-world Signals}
In this subsection, we perform dynamic sensor placement for real data to validate the proposed method for practical observations.

\subsubsection{Setup}
We use the global sea surface temperature data \cite{rayner_global_2003} during 2016--2021.
This dataset is composed of snapshots recorded every month from 2016 to 2021. The first five months are used for training ($T_{\text{train}}=5$), while the remaining months are used for testing ($T_{\text{test}}=55$). The initial data slot $\mb{X}_0$ of the dictionary learning is set to the same scenario as in the synthetic experiments. Temperature is sampled at intersections of 1-degree latitude-longitude grids. For simplicity, we clip snapshots of the west coast of the U.S. and randomly sample 100 nodes for the experiments. Then, we construct a $5$-nearest neighbor graph. 
The numbers of time slots and sensors are set to $D=5$ and $K=10$, respectively.
The parameters of the dictionary learning are set to $(\gamma,\eta,\mu)=(10^{-3}, 1, 1)$, and $P=1$ is used for the sensor movable area. 
We design the filter matrix $\mb{G}$ the same as \eqref{eq:graph_filter} in Section~\ref{sec:experiments}-\ref{subsec:synthetic}.
We evaluate the averaged MSEs and compare them with the above-mentioned two benchmark methods Static1 and 2.

\begin{figure}[t!]
    \centering
    \includegraphics[width = 0.8\linewidth]{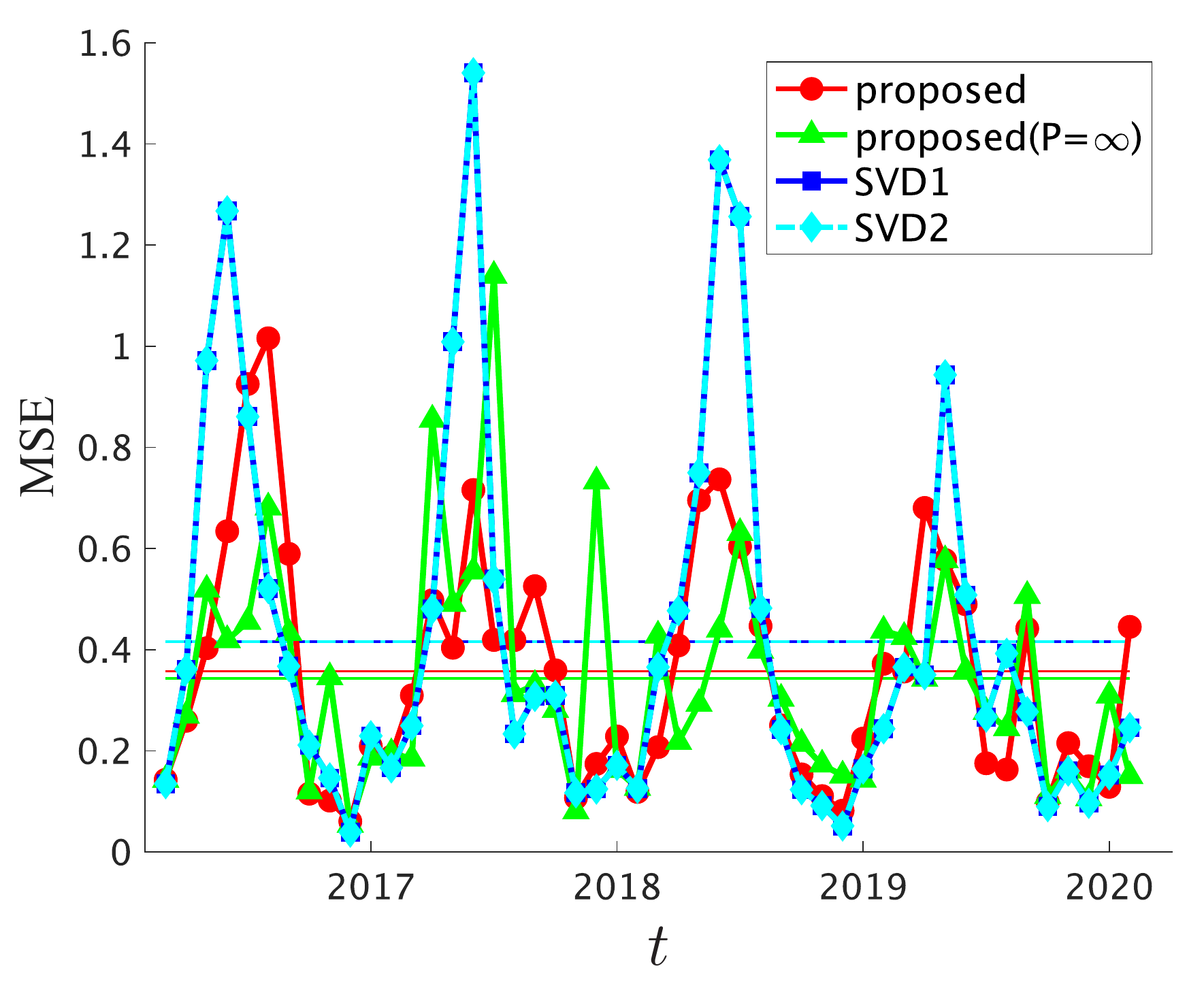}
    \caption{Comparison of reconstruction MSEs of sea surface temperature data. The average MSE for each of the methods is plotted as a horizontal line.}
    \label{fig:mse_sst}
\end{figure}

\begin{figure*}[t!]
    \centering
    \includegraphics[keepaspectratio, width=0.9\linewidth]{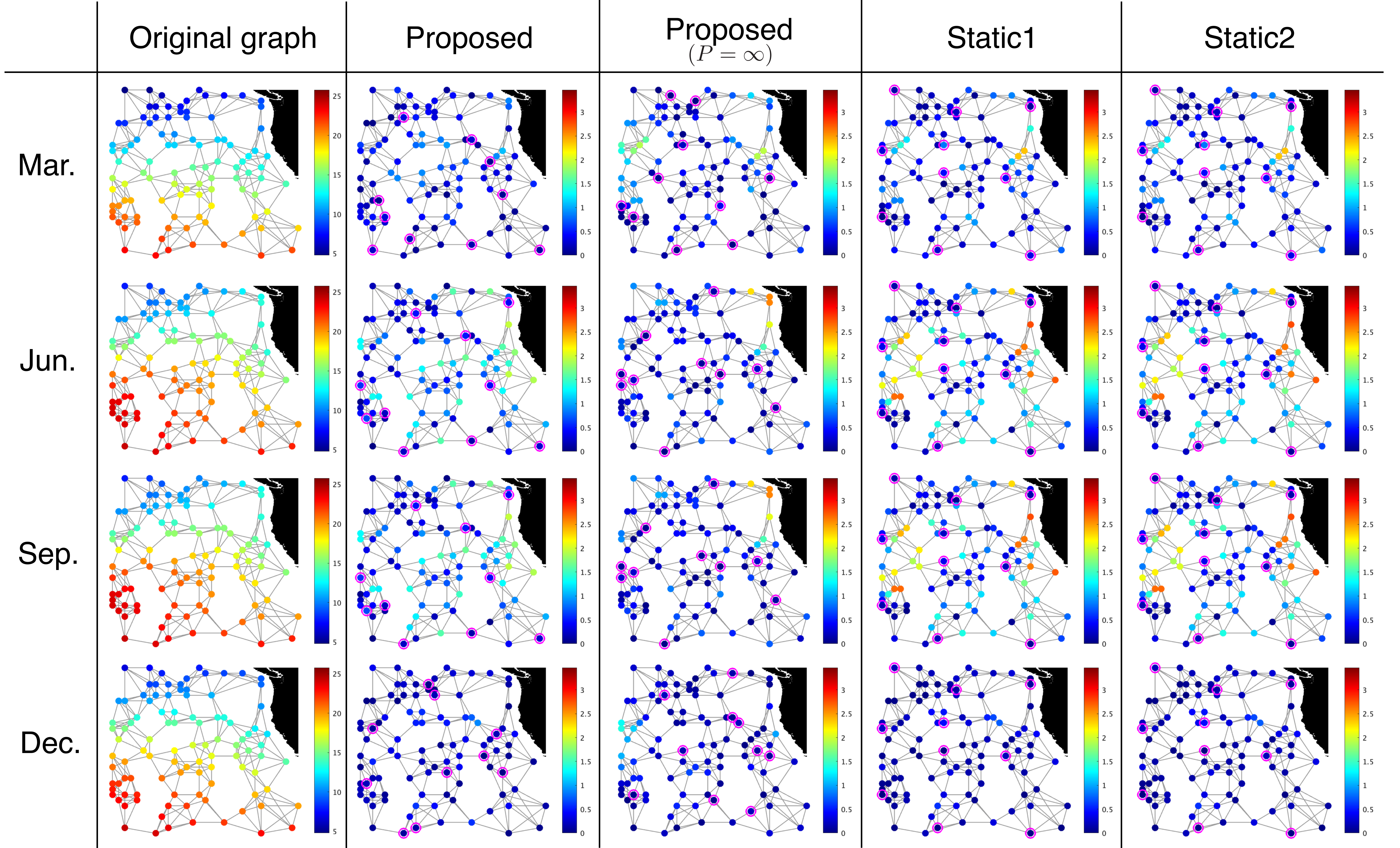}
    \caption{Reconstructed errors of sea surface temperature data. Circled nodes represent sensor positions.}
    \label{fig:reconstruct_sst}
\end{figure*}

\begin{figure}[t!]
    \centering
    \includegraphics[keepaspectratio, width=0.9\linewidth]{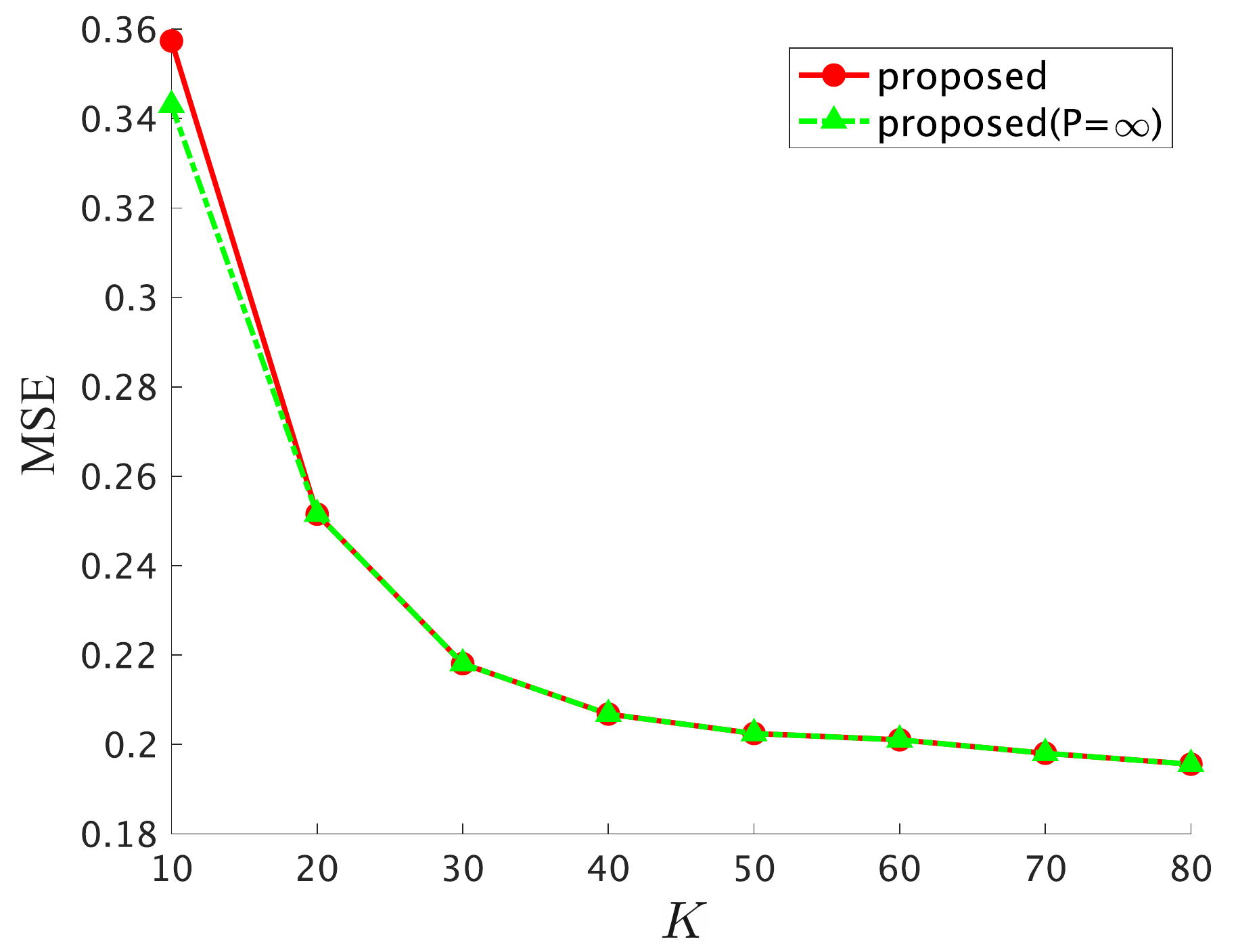}
    \caption{Comparison of the averaged MSE of the proposed method with different numbers of sensors $K$.}
    \label{fig:dif_K}
\end{figure}

\subsubsection{Results}
Fig.~\ref{fig:mse_sst} shows the MSE comparison.
As with the experiment for the synthetic data, the proposed dynamic sensor placement presents lower MSEs than the SVD-based methods for most $t$s.
The MSEs of Static1 and Static2 are almost the same.

The reconstruction errors between the original and reconstructed graph signals are visualized in Fig.~\ref{fig:reconstruct_sst}, along with the selected sensor positions.
The proposed methods place many sensors in the southern area in March and June,  while the positions are more uniformly distributed in September and December.
This may reflect the fact that sea surface temperatures in the northern region do not change much from December to April, while those in the southern region change significantly, as shown in Fig.~\ref{fig:reconstruct_sst}. 
Therefore, in order to capture such changes in sea surface temperatures, the sensors gradually moved to the south after December. 

We also compare the averaged MSE of the proposed method with varying numbers of sensors (different values of $K$). Fig.~\ref{fig:dif_K} shows the experimental results with different $K$. As expected, the proposed method tends to have a smaller average MSE for a larger $K$.

\section{CONCLUSION}
In this paper, we propose a dynamic sensor placement method based on graph sampling theory.
We sequentially learn the dictionary from a time series of observed graph signals by utilizing sparse coding. Using the dictionary, we dynamically determine the sensor placement at every time instance such that the non-observed graph signal values can be best recovered from those of the observed (selected) nodes.
In experiments, we demonstrate that the proposed method outperforms existing static sensor placement methods in synthetic and real datasets.

\bibliographystyle{IEEEtran}
\bibliography{IEEEabrv, ref_saki_nomura}

\end{document}